\documentclass[a4paper,1p]{elsarticle}
\usepackage{graphicx}
\usepackage{amsmath}
\usepackage{amssymb}
\journal{Physica A}
\begin{document}
\begin{frontmatter}
\title{%
A contribution to the systematics of stochastic volatility models
}%%
\author[ada]{%
Franti\v{s}ek Slanina
}
\ead{slanina@fzu.cz
}
\address[ada]{%
Institute of Physics,
 Academy of Sciences of the Czech Republic,\\
 Na~Slovance~2, CZ-18221~Praha,
Czech Republic\\
and Center for theoretical study, Jilsk\'a 1, Praha, Czech Republic
}
\begin{abstract}
We compare systematically several classes of stochastic volatility
models of stock market fluctuations. We show 
that the long-time return distribution is either 
Gaussian or develops a power-law tail, while the 
short-time return distribution has generically a 
stretched-exponential form, but can assume also an algebraic
decay, in the family of models which 
we call ``GARCH''-type. The intermediate regime is found in the exponential
Ornstein-Uhlenbeck process. 
We calculate also the decay of the autocorrelation function of volatility.
\end{abstract}


\end{frontmatter}

\section{Introduction}

The modelling based on stochastic volatility
\cite{lewis_00,cis_fer_mon_nic_07} is a natural 
way how to implement non-Markovian property of certain stochastic
processes into the machinery of stochastic differential equations,
which are Markovian by definition. The idea is old, well-known
 and straightforward
and amounts postulating an auxiliary process (or processes), which
together with our process of interest form a multicomponent process,
which is Markovian, while each of the components itself is not. The
auxiliary processes may have a direct physical interpretation or not,
depending on the situation.

When modelling the stock-market fluctuations, the process of interest
is the price (which is directly measurable) and the auxiliary process
is the instantaneous volatility, or some function of it, which is,
however, purely hypothetical and in principle not measurable. We can
estimate its properties from the volatility measured within certain
time interval, but nevertheless, the instantaneous volatility remains
an elusive ghost. Supposing it is subject to a process described by a
stochastic differential equation, we can solve the equation pair for
price and volatility and deduce for example the return distribution,
autocorrelations etc.

Since this methodology is quite general, it is of interest not only
for economists but can be useful in various purely physical
situations. One of such examples might be the stochastic resonance under the 
influence of coloured noise. In this paper we present some generalisations of
previously known results, stressing the generic and the special
features of several classes of stochastic volatility models.

\section{Systematics}

Quite general ensemble of stochastic volatility models is described by a
pair of stochastic differential equations (in Ito convention) with algebraic coefficients
\begin{equation}
\begin{split}
&\mathrm{d}X_t=S_t^\gamma\,\mathrm{d}W_{1t}\\
&\mathrm{d}S_t=a\,(\sigma-S_t)\,S_t^\alpha\,\mathrm{d}t
+g\,S_t^\beta\,\mathrm{d}W_{2t}\;.
\label{eq:sv-general}
\end{split}
\end{equation}
We denote $W_{1t}$ and $W_{2t}$ two independent Wiener processes with
unit diffusion constant,
$[\mathrm{d}W_{1t}]^2=[\mathrm{d}W_{2t}]^2=\mathrm{d}t$. In this
article we 
disregard the possible dependence of the processes $W_{1t}$ and
$W_{2t}$, which was investigated e. g. in
\cite{bou_mat_pot_01,mas_per_02,per_mas_bou_03,per_mas_03,per_mas_ane_04}
in the context of the leverage effect. We also omit the possibility
that either $W_{1t}$ or $W_{2t}$ 
is a L\'evy noise \cite{bar_she_01}, because that would
lead us to a completely different area.

The process $X_t$ describes the quantity of interest, which is observable
and if we remain in the field of econophysics, it is typically the logarithm of
the price of a
commodity or a security. We shall call it simply the price and allow
both positive and negative values for it.
$S_t$ is the auxiliary process, representing a hypothetical quantity
which will be called volatility. Contrary to the true volatility,
which is (whatever its definition might be) genuinely two-time quantity, like the return, 
the ``volatility'' $S_t$ is an instantaneous, one-time stochastic
variable. This does not mean that it is completely inaccessible to
observations, but implies that its relation to observations is not a
direct one and poses a separate problem, not to be investigated here.
We consider necessary to recall these trivial observations, as
they are often neglected in the econophysics literature.

Some of the combinations of the exponents bear their usual names. We make an overview in the
following table.
\begin{center}
\begin{tabular}{|lll|r|}
\hline
$\alpha$&$\beta$&$\gamma$&~~name or acronym of the model~~\\
\hline
$0$&$0$&$1$&Stein-Stein\\
$0$&$0$&$1/2$&''Ornstein-Uhlenbeck'' or ``OU''\\
$0$&$1/2$&$1/2$&Heston\\
$0$&$1$&$1/2$&''GARCH''\\
$1$&$1$&$1/2$&''geometric OU''\\
$0$&$3/2$&$1/2$&$3/2$-model\\
\hline
\end{tabular}
\end{center}
(The names in quotation marks ``~~'' are somewhat abuses of notation,
because properly speaking they are already in use in more or
less different sense.) 
Besides these models which contain only coefficients depending on $S$
as a power, we shall investigate also the exponential
Ornstein-Uhlenbeck process (acronym expOU) \cite{cis_fer_mon_nic_07,per_mas_03,per_mas_ane_04,scott_87,lebaron_01,mas_per_05,eis_per_mas_06,per_sir_mas_08,bor_caz_mon_nic_08,mas_per_07},  which in its simplest form
is described by the pair of equations
\begin{equation}
\begin{split}
&\mathrm{d}X_t=\mathrm{e}^{\sigma+S_t}\,\mathrm{d}W_{1t}\\
&\mathrm{d}S_t=-a\,S_t\,\mathrm{d}t
+g\,\mathrm{d}W_{2t}\;.
\label{eq:sv-expou}
\end{split}
\end{equation}
Again, $X_t$ is the logarithm of price and $S_t$ the hypothetical
hidden variable. The instantaneous volatility in this case is $\mathrm{e}^{2(\sigma + S_t)}$.

If we were interested only in the auxiliary process $S_t$, we would
need to solve only the second of the equations
(\ref{eq:sv-general}). Processes of this type were solved in many
contexts and are considerably easier compared to solving the coupled
pair $(X_t,S_t)$. The older works relevant to our study are
e. g. \cite{bar_ciu_spa_93} and \cite{ciu_pas_spa_93}. In the former, 
the special case  $\alpha=0$,
$\beta=1$, $\sigma=0$ is solved exactly using the Lie-group
technique. The latter treats approximately the
difficult case $\alpha=1$, $\beta=1/2$ which we avoid completely, as
we never found it used in the stochastic volatility context.

Let us make a very short historical remark at the end of this section.
The study of stochastic volatility models in economy can
be traced back to the pioneering works of Vasicek  \cite{vasicek_77} and Cox,
Ingersoll and Ross  \cite{cox_ing_ros_85} on interest rates. In fact, the same problem was investigated, on the level of
Fokker-Planck equation, much earlier by Feller \cite{feller_51}. Hull
and White  \cite{hul_whi_90} formulated the problem on a more general basis.
These results were adapted for the modelling of price fluctuations by
Scott  \cite{scott_87}, Hull and White  \cite{hul_whi_87}, Stein
and Stein \cite{ste_ste_91}. The Stein-Stein model was thoroughly
studied \cite{ste_ste_91,bal_rom_94,sch_zhu_99}, although it exhibits
some unnatural features, namely negative volatilities. 
These are absent in the Heston model \cite{heston_93}, which became very popular
\cite{bal_rom_94,dra_yak_02,dra_yak_02b,sil_yak_03,sil_pra_yak_04,vic_tol_lei_cat_04,sil_yak_06}.
Of course, the potential of stochastic volatility models was used for
better prediction of investment risks, see e. g.
\cite{hul_whi_87,scott_87,heston_93,bal_rom_94,jiz_kle_hae_07,per_sir_mas_08}. 
For further variants of the stochastic volatility models used in
econophysics see e. g. references
\cite{pal_per_mon_mas_04,ric_sab_04,ant_rie_05,labordere_05,labordere_05a,bon_val_spa_06,bon_val_spa_05,val_spa_bon_06,villaroel_06,spa_val_08}.

\section{Long-time properties}

The properties of the process (\ref{eq:sv-general}) can be studied
directly by the method of moments. This way we can easily find
indirect evidence on the asymptotic distribution of price
change. Indeed, we shall find that in the most important cases the whole
set of asymptotic moments can be calculated and that they fall into
two groups. Either the moments imply the Gaussian distribution of
returns, or
some of them diverge, indicating power-law tail in the return distribution. 

On the other hand, the method of moments has also serious
disadvantages. Most importantly, it does not allow an easy
implementation of the requirement of positive volatility. For $\gamma=1$ this problem does
not occur, because the ``physical'' volatility is the square $S_t^2$,
but for $\gamma=1/2$ we must consider it.
When solving the
set of Fokker-Planck equations, it is imposed as a boundary
condition. On the level of moments we do not have such
possibility. Therefore, we either use the fact that the process itself
prohibits trajectories connecting positive and negative volatilities
(this is the case of $\beta>0$) or assume that the events with
negative volatilities are rare enough to be neglected. Typically it
means that $\sigma\gg g/\sqrt{a}$. This is how we shall treat the case
$\beta=0$.

The second disadvantage resides in the fact that we are unable to
calculate the moments for all values of the exponents $\alpha$,
$\beta$, and $\gamma$. If all of them are rational, we can bring them
to a common denominator, say, $q$, and calculate moments which are
integer multiples of $1/q$. For large $q$ it is possible but
difficult. However, for irrational and incommensurate values of the exponents 
this way is unfeasible.

The general moment is defined as $\mu_{m,n}(t)=\langle
X_t^m\,S_t^n\rangle$ and its time evolution according to the process
(\ref{eq:sv-general}) is contained in the equation
\begin{equation}
\begin{split}
\frac{\mathrm{d}}{\mathrm{d} t}\mu_{m,n}(t&)
=na\,\Big(\sigma\,\mu_{m,n+\alpha-1}(t)
-\mu_{m,n+\alpha}(t)\Big)+\\
&+\frac{1}{2}m(m-1)\,\mu_{m-2,n+2\gamma}(t)
+\frac{g^2}{2}n(n-1)\,\mu_{m,n-2+2\beta}(t)\;.
\end{split}
\label{eq:sv-moments-equation}
\end{equation}
We do not have a solution in a closed form in the general case, but we can
investigate each model separately. We shall follow essentially the
order in which the models are listed in the table above.

We start with the Stein-Stein model
 and proceed from small values of
$m$ and $n$ to higher ones. For $m=0$ the equation is
\begin{equation}
\dot{\mu}_{0,n}+na\,\mu_{0,n}=na\sigma\,\mu_{0,n-1}+\frac{g^2}{2}n(n-1)\,\mu_{0,n-2}
\label{eq:sv-stein-stein-moments-zero-n-equation}
\end{equation}
and we can see that knowing $\mu_{0,n-1}$ and $\mu_{0,n-2}$ we obtain
$\mu_{0,n}$ by trivial integration. The starting point
$\mu_{0,0}(t)=\langle 1\rangle=1$ is obvious. The next two moments are
\begin{equation}
\begin{split}
&\mu_{0,1}(t)=\sigma+\big(\mu_{0,1}(0)-\sigma\big)\,\mathrm{e}^{-at}\\
&\mu_{0,2}(t)=\sigma^2+\frac{g^2}{2a}
+\Big[\mu_{0,2}(0)-\Big(\sigma^2+\frac{g^2}{2a}\Big)\Big]\,\mathrm{e}^{-2at}+\\
&\qquad\qquad\qquad\;\;\;\;\;\;
+2\sigma\,\big(\mu_{0,1}(0)-\sigma\big)\,\Big(\mathrm{e}^{-at}-\mathrm{e}^{-2at}\Big) 
\end{split}
\label{eq:sv-stein-stein-moments-zero-onetwo-solution}
\end{equation}
and we can continue further as far as we please. The generic feature
of the time evolution of moments, seen in
(\ref{eq:sv-stein-stein-moments-zero-onetwo-solution}) is an
exponential relaxation to a stationary value. Setting
$\dot{\mu}_{0,n}=0$ in
(\ref{eq:sv-stein-stein-moments-zero-n-equation})  
we obtain a chain of equations for the stationary values of the
moments $\mu_{0,n}(\infty)$. First two of them,
$\mu_{0,1}(\infty)=\sigma$,
$\mu_{0,2}(\infty)=\sigma^2+\frac{g^2}{2a}$
 can be seen in the formulae
 (\ref{eq:sv-stein-stein-moments-zero-onetwo-solution}), and the
 general result is
\begin{equation}
\mu_{0,n}(\infty)=
\sum_{k=0}^{\lfloor n/2\rfloor}
\binom{n}{2k}\,\frac{(2k)!}{2^k\,k!}\,\Big(\frac{g^2}{2a}\Big)^k
\,\sigma^{n-2k}
\label{eq:sv-stein-stein-moments-zero-threefour-infty}
\end{equation}
confirming that the stationary distribution of $S_t$ only is Gaussian with average $\sigma$ and
variance $g^2/(2a)$, as expected. (The fact that the distribution is
Gaussian follows immediately from the comparison with the formula for
$2k$-th moment of a Gaussian distribution with variance $s$, 
which is $[(2k)!/(2^k\,k!)]\,s^k$.)

The moments with $m>0$ are just a bit more complicated. We need to
investigate only even $m$, because all moments for odd $m$ are
zero. Integrating the equation
(\ref{eq:sv-moments-equation}) we can get the moments for
higher and higher $m$ systematically. The most interesting part of the
result is the behaviour of the moments for large times. 
For example, we find, using
(\ref{eq:sv-stein-stein-moments-zero-onetwo-solution}),
that  
\begin{equation}
\mu_{2,0}(t)=\mu_{2,0}(0)+\int_0^t\mu_{0,2}(\tau)\,\mathrm{d}\tau\simeq
\Big(\sigma^2+\frac{g^2}{2a}\Big)\,t\text{~,~~~as~~~}t\to\infty\;.
\end{equation}
Generally, it is possible to show that
$\big(\frac{\mathrm{d}}{\mathrm{d} t}\big)^l\mu_{2l,n}(t)$ has a
finite limit when $t\to\infty$ and after some easy algebra we arrive
at a formula summarising the long-time behaviour of all moments
\begin{equation}
\lim_{t\to\infty}\,t^{-l}\,\mu_{2l,n}(t)
=\frac{(2l)!}{2^l\,l!}\,\Big(\sigma^2+\frac{g^2}{2a}\Big)^l\,\mu_{0,n}(\infty)\;.
\label{eq:sv-stein-stein-moments-longtime}
\end{equation}
This result 
indicates that for long times the logarithmic returns have Gaussian
distribution
 with variance $\big(\sigma^2+\frac{g^2}{2a}\big)\,t$.

Analogous steps ought to be taken when we want to find the moments in
the case of the ``OU'' model. Note that the moments including only the
process $S_t$, i. e. those with $m=0$, are identical to the
Stein-Stein model.
 Therefore, the formulae
(\ref{eq:sv-stein-stein-moments-zero-onetwo-solution}) and
(\ref{eq:sv-stein-stein-moments-zero-threefour-infty})
remain in force. The long-time asymptotics of the remaining moments is
obtained in the same way as Eq. 
(\ref{eq:sv-stein-stein-moments-longtime}). This time, the result is 
\begin{equation}
\lim_{t\to\infty}\,t^{-l}\,\mu_{2l,n}(t)
=\frac{(2l)!}{2^l\,l!}\,\sigma^l\,\mu_{0,n}(\infty)
\label{eq:sv-ou-moments-longtime}
\end{equation}
and again suggests Gaussian distribution of the returns.

The next in line is the Heston model.
 The moments evolve according to
the set of equations
\begin{equation}
\dot{\mu}_{m,n}+na\,\mu_{m,n}=na\big(\sigma+\frac{g^2}{2a}(n-1)\big)\,\mu_{m,n-1}
+\frac{1}{2}m(m-1)\,\mu_{m-2,n+1}
\label{eq:sv-heston-moments-equation}
\end{equation}
and solving them for the lowest non-trivial moments we get
\begin{equation}
\begin{split}
&\mu_{0,1}(t)=\sigma+\big(\mu_{0,1}(0)-\sigma\big)\,\mathrm{e}^{-at}\\
&\mu_{2,0}(t)=\mu_{2,0}(0)+\sigma\,t+\frac{1}{a}\big(\mu_{0,1}(0)-\sigma\big)\big(1-\mathrm{e}^{-at}\big)\;.
\end{split}
\label{eq:sv-heston-moments-solution}
\end{equation}
The long-time behaviour is again obtained assuming that the
derivatives $\big(\frac{\mathrm{d}}{\mathrm{d} t}\big)^l\mu_{2l,n}(t)$
have finite limit for $t\to\infty$ and checking a posteriori that this
assumption does not contain any contradiction. We can do even better
than we could for Stein-Stein
 and ``OU'' models. The limits of
the moments are expressed by the closed formula
\begin{equation}
\lim_{t\to\infty}\,t^{-l}\,\mu_{2l,n}(t)
=\frac{(2l)!}{2^l\,l!}\,\sigma^l
\,
\Big(\frac{g^2}{2a}\Big)^n
\,
\frac{\Gamma\big(\frac{2a\sigma}{g^2}+n\big)}{\Gamma\big(\frac{2a\sigma}{g^2}\big)}
\label{eq:sv-heston-moments-longtime}
\end{equation}
showing again without any doubt that the long-time distribution of
returns is Gaussian.

The same procedure is effective also for the ``GARCH'' model, where
the moments evolve according to
\begin{equation}
\dot{\mu}_{m,n}+na\,\Big(1-\frac{g^2}{2a}(n-1)\Big)\mu_{m,n}=na\,\sigma\,\mu_{m,n-1}
+\frac{1}{2}m(m-1)\,\mu_{m-2,n+1}\;.
\label{eq:sv-garch-moments-equation}
\end{equation}
The evolution of the lowest moments $\mu_{0,1}(t)$ and $\mu_{2,0}(t)$
follows exactly the same 
law (\ref{eq:sv-heston-moments-solution}) as in the case of the Heston
model. Moreover, we have
\begin{equation}
\begin{split}
\mu_{0,2}(t)=\frac{2a\,\sigma^2}{2a-g^2}
+&\Big(\mu_{0,2}(0)-\frac{2a\,\sigma^2}{2a-g^2}\Big)\,\mathrm{e}^{-(2a-g^2)t}+\\
+&2a\,\sigma\,\frac{\mu_{0,1}-\sigma}{a-g^2}\,\mathrm{e}^{-at}
\big(1-\mathrm{e}^{-(a-g^2)t}\big)\;.
\end{split}
\label{eq:sv-garch-moments-solution}
\end{equation}
We must be more careful here when we compute the asymptotics of the
moments. Some of them diverge exponentially, rather than
algebraically, with $t$, becoming effectively infinite. We find that
the limit
\begin{equation}
\lim_{t\to\infty}\,t^{-l}\,\mu_{2l,n}(t)
=\frac{(2l)!}{2^l\,l!}\,\sigma^{l+n}
\,
\Big[\prod_{j=1}^{n-1}\big(1-j\frac{g^2}{2a}\big)\Big]^{-1}
\label{eq:sv-garch-moments-longtime}
\end{equation}
is finite as long as $l+n<1+\frac{2a}{g^2}$. Otherwise the moment
should be considered diverging. The interpretation of this result is,
that the distribution develops power-law tails. It should hold both
for the marginal distribution of returns, described by the moments
with $n=0$, and the marginal distribution of volatility, corresponding
to the set of moments with $l=0$. 
Hence, the tail exponent of the return distribution is
\begin{equation}
\tau = 3+\frac{4a}{g^2}\;.
\end{equation}
Therefore, the exponent is always bounded from below by $\tau>3$, a result
which is well consistent with the empirical findings
\cite{go_ple_am_me_sta_99,ple_gop_am_mey_sta_99}. Note also that the
empirical volatility distribution
\cite{ci_li_me_pe_sta_97,li_go_ci_me_pe_sta_99,bou_pot_03,mic_bon_lil_man_02,cis_fer_mon_nic_07}
shows log-normal central part, 
followed by a power-law tail, which is consistent with the power-law
tail in the distribution of the hypothetical volatility $S_t$, as is
evident from (\ref{eq:sv-garch-moments-longtime}).  All of that
suggests that the ``GARCH'' member of the stochastic-volatility family
of stock market models is close to reality.

The next two models in the list are the ``geometric OU''  and the
$3/2$ models. Unfortunately, the method of moments does not lead to a
recursive chain which would allow computing all the moments. The only
way forward is to solve directly the corresponding Fokker-Planck
equations. We shall see later how we can bypass it in the short-time
approximation. 

The treatment of the expOU model using the method of moments is
slightly more involved, as it requires expansion of the exponential in
powers and later resummation of a series back, to get a compact
result. The form of the equations (\ref{eq:sv-expou}) ensures that the
odd moments are zero, $\mu_{m,2k-1}=\mu_{2l-1,n}=0$, if the initial
condition is $X_0=S_0=0$. In that case, we find, for the even moments
\begin{equation}
\lim_{t\to\infty}\,t^{-l}\,\mu_{2l,2k}(t)
=\frac{(2l)!}{2^l\,l!}\,
\exp\Big[l\big(2\sigma+\frac{g^2}{a}\big)\Big]
\,
\frac{(2k)!}{2^k\,k!}\,
\Big(\frac{g^2}{2a}\Big)^k\;.
\label{eq:sv-expouou-moments-longtime}
\end{equation}
This type of formula is already familiar to us. The interpretation is
that at long times the logarithm of price is Gaussian-distributed with variance
$t\exp(2\sigma+g^2/a)$.

\section{Autocorrelations}

The autocorrelation function of returns
\begin{equation}
C_{\mathrm{ret}\,q}(\Delta t;\delta t)=\langle
|(X_{t+\delta t}-X_t)(X_{t+\Delta t+\delta t}-X_{\Delta t})|^q\rangle
\label{eq:sv-autocorrelation-ret}
\end{equation}
is a four-time quantity, not easy to calculate by the method of
moments. Instead, we shall investigate a simpler quantity, which is
supposed to share relevant properties with the autocorrelation function
(\ref{eq:sv-autocorrelation-ret}). Indeed, for short time intervals
$\delta t$ the absolute increments of price are proportional to
$S_t^\gamma$ and we can consider the autocorrelation function
\begin{equation}
C_q(\Delta t)=\langle
|S_{t}\,S_{t+\Delta t}|^{q\gamma}\rangle
\label{eq:sv-autocorrelation-vol}
\end{equation}
as an approximation for $C_{\mathrm{ret}\,q}$ of $\delta t\to 0$.
In stationary state it is a function of one time argument only and we
obtain for it exactly the same differential equations as we have had
for the moments. To this end we define a slightly more general
autocorrelation
\begin{equation}
C_{u,v}(\Delta t)=\langle
S_{t}^u\,S_{t+\Delta t}^v\rangle
\label{eq:sv-autocorrelation-vol-gen}
\end{equation}
and obtain the following equation for it
\begin{equation}
\frac{\mathrm{d}C_{u,v}(\Delta t)}{\mathrm{d}\Delta t}=
-va\big(C_{u,v+\alpha}(\Delta t)-\sigma\,C_{u,v-1+\alpha}(\Delta
t)\big)
+\frac{g^2}{2}v(v-1)\,C_{u,v-2+2\beta}(\Delta t)\;.
\end{equation}
We can see that the autocorrelation function with $v=1$ does not
depend on the value of $\beta$. Especially, all the models examined in
detail above have $\alpha=0$ and the solution for all of them is
\begin{equation}
C_{u,1}(\Delta t)=
     \sigma\,\mu_{0,u}(\infty)+\big(\mu_{0,u+1}(\infty)-\sigma\,\mu_{0,u}(\infty)\big)\,
     \mathrm{e}^{-a\Delta t}\;.
\label{eq:sv-autocorrelation-vol-1}
\end{equation}

For $v>1$ the value of $\beta$ is relevant. We could proceed
iteratively finding the correlation function for $v=2$, then for $v=3$
etc. However, it is possible to express the solution as a linear
combination of exponentials and then solve the set of equations for
the coefficients. Thus we find the full solution in relatively compact form.
In the case $\alpha=\beta=0$ (i. e. Stein-Stein and
``OU'' models), we get
\begin{equation}
C_{u,v}(\Delta t)=\sum_{m=0}^{v}\binom{v}{m}
     \Bigg[
\sum_{n=0}^{\lfloor\frac{m}{2}\rfloor}
\frac{(2n)!}{2^n\,n!}\binom{m}{2n}\sigma^{m-2n}\Big(\frac{g^2}{2a}\Big)^n\Bigg]
\,f_{v-m}\,\mathrm{e}^{-(v-m)a\,\Delta t}
\label{eq:sv-autocorrelation-vol-steinstein}
\end{equation}
where $f_l$ are constants determined from the initial conditions
$C_{u,v}(0)=\mu_{0,u+v}(\infty)$. 

Similarly, for the Heston model ($\beta=1/2$) we have
\begin{equation}
C_{u,v}(\Delta t)=\sum_{l=0}^{v}\binom{v}{l}
\frac{\Gamma\Big(\frac{2a\sigma}{g^2}+v\Big)}{\Gamma\Big(\frac{2a\sigma}{g^2}\Big)}
\Big(\frac{g^2}{2a}\Big)^v\,f_l\,\mathrm{e}^{-la\,\Delta t}\;.
\label{eq:sv-autocorrelation-vol-heston}
\end{equation}

The case $\beta=1$ (``GARCH'') is more complicated. We could also
try to find general expressions like (\ref{eq:sv-autocorrelation-vol-steinstein})
and (\ref{eq:sv-autocorrelation-vol-heston}), but the decay rates of
the exponentials are not ordered in arithmetic sequence
$\{la\}_{l=0}^\infty$, but form a two-parametric set
$\{l_1\,a-l_2\,g^2/2\;|\;l_1,l_2=0,1,2\ldots\}$. Some of these rates
are negative, which indicates that  some of the correlation functions
do not exist in stationary state. This feature stems from the
divergence of higher moments of the volatility in stationary
regime. However, the general conclusion holds, that if the stationary
autocorrelation function does exist, it decays as a combination of
exponentials. 

We can conclude that 
the common feature of all stochastic volatility models investigated
here is the exponential decay of 
autocorrelations. This is natural, because the process $S_t$ has
Markov property. 
For practical use, we note that the multiple timescales found in the
expOU model  \cite{lebaron_01,mas_per_05} can successfully reproduce
the slow decay of autocorrelations found in empirical data.
Special cases when
the autocorrelation does decay as a power law are examined e. g. in
\cite{suz_kan_tak_82}, but these cases do not seem to have much
relevance for stock-market modelling.

\section{Short-time properties}

From the point of view of return distributions,
the long-time properties of the stochastic volatility models discussed
so far do not 
exhibit much  diversity. The return distribution is mostly
Gaussian, with the exception of the ``GARCH''
model, where power-law tails develop. At short times, however, more
variable results are found. Indeed, the empirical return distribution tend all
to a Gaussian for large enough times, so it is the range of medium
time distances at which the interesting fat tails are observed. So, it
is well sensible to discuss the special features of the stochastic
volatility models appearing at short times, i. e. at times shorter
than those at which Gaussian limit is reached. We have seen that the
typical decay time of autocorrelations is $1/a$. We expect that this
is also the time when the Gaussian distribution starts to
prevail. Therefore, we shall investigate the return distributions for
time distances $\Delta t\ll a^{-1}$.
This kind of approach is sometimes related \cite{bir_ros_07} 
to the
Born-Oppenheimer approximation
 used in quantum chemistry.

\subsection{Volatility distribution}

 The ingredient needed for these investigations will be
the stationary distribution for the volatility process $S_t$. It is
easy to find it, because  the
instantaneous volatility influences the price process $X_t$, but is
not affected by it, so we can find the distribution for $S_t$
separately. We must solve the Fokker-Planck equation
\begin{equation}
\begin{split}
\frac{\partial}{\partial t}P_{S,t}(s)
=&\,a\frac{\partial}{\partial s}\big[(s-\sigma)s^\alpha\,P_{S,t}(s)\big]+\\
&+\frac{g^2}{2}\frac{\partial^2}{\partial s^2}
\big[s^{2\beta}\,P_{S,t}(s)\big]\;.
\end{split}
\label{eq:fokkerplanck-marginal}
\end{equation}
in the limit $t\to\infty$. Contrary to the method of moments, we can
obtain the solution generally for any value of the parameters $\alpha$
and $\beta$. It is also easy to implement the reflective boundary
condition at $S_t=0$, enforcing the positivity of the volatility
simply by multiplying the solution of (\ref{eq:fokkerplanck-marginal})
by the Heaviside function $\theta(s)$ and recomputing the
normalisation factor so that $\int_0^\infty P_S(s)\mathrm{d}s=1$. This
will work for all solutions with zero total current. 

 We can get easily the generic solution with zero current in the form
\begin{equation}
P_S(s)=\mathcal{N}\,\theta(s)\,s^{-2\beta}\,\exp\Big(
-\frac{2a}{g^2}\,\frac{s^{2+\alpha-2\beta}}{2+\alpha-2\beta}
+\frac{2a\,\sigma}{g^2}\,\frac{s^{1+\alpha-2\beta}}{1+\alpha-2\beta}
\Big)
\label{eq:fokkerplanck-marginal-solution-general}
\end{equation}
where $\mathcal{N}$ is the appropriate normalisation constant. The
Gaussian distribution is included here, for the combination of
parameters giving $\alpha-2\beta=0$.

 There
are two exceptional cases, marked by the vanishing denominators in the
exponent in
Eq. (\ref{eq:fokkerplanck-marginal-solution-general}). When
$1+\alpha-2\beta=0$ we have
\begin{equation}
P_S(s)=\mathcal{N}\,\theta(s)\,s^{-2\beta+{2a\,\sigma/g^2}}\,\exp\Big(
-\frac{2a}{g^2}\,s\Big)\;.
\label{eq:fokkerplanck-marginal-solution-hestontype}
\end{equation}
We shall call this case ``Heston type'', as the Heston model belongs to
it. The second exceptional case occurs when $2+\alpha-2\beta=0$. The
probability density for the volatility is then
\begin{equation}
P_S(s)=\mathcal{N}\,\theta(s)\,s^{-2\beta-{2a/g^2}}\,\exp\Big(
-\frac{2a\,\sigma}{g^2\,s}\Big)\;.
\label{eq:fokkerplanck-marginal-solution-garchtype}
\end{equation}
This case will be named ``GARCH-type'', because it includes the
``GARCH'' model studied above.

\subsection{Return distribution}

 We can neglect fluctuations of
the volatility during time delays $\Delta t\ll \alpha^{-1}$ and consider
the volatility fixed with distribution given by 
(\ref{eq:fokkerplanck-marginal-solution-general}), or, in special
cases, by 
(\ref{eq:fokkerplanck-marginal-solution-general}) or 
(\ref{eq:fokkerplanck-marginal-solution-hestontype}). Thus, our
starting formula will be
\begin{equation}
P_{\Delta X}(\Delta x;\Delta t)=\int_0^\infty
s^{-\gamma}\,\exp\Big(\frac{(\Delta x)^2}{2\Delta
  t\,s^{2\gamma}}\Big)P_S(s)\frac{\mathrm{d}s}{\sqrt{2\pi\,\Delta
    t\,}}\;.
\label{eq:short-time-integral}
\end{equation}
Contrary to the method of moments, we can proceed directly with the
calculation for arbitrary values of the parameters $\alpha$, $\beta$,
$\gamma$. We shall explore this freedom for $\alpha$ and $\beta$, bit
for the parameter $\gamma$ we use only the most studied values $1/2$
and $1$. The general strategy will be to use a suitable change of
variables to perform the integral (\ref{eq:short-time-integral}) by
the saddle-point method.

\subsubsection{$\gamma=1/2$}

In the general case (\ref{eq:fokkerplanck-marginal-solution-general})
we define 
\begin{equation}
\begin{split}
\xi=&\,\Big[\frac{(\Delta x)^2}{\Delta
    t}\Big]^{\frac{2+\alpha-2\beta}{3+\alpha-2\beta}}
\;\Big[\frac{4a}{g^2(2+\alpha-2\beta)}\Big]^{\frac{1}{3+\alpha-2\beta}}\\
\eta=&\,\sigma\,\Big[\frac{g^2}{4a}\,\frac{(\Delta x)^2}{\Delta
    t}\Big]^{-\frac{1}{3+\alpha-2\beta}}
\frac{(2+\alpha-2\beta)^{\frac{2+\alpha-2\beta}{3+\alpha-2\beta}}}{1+\alpha-2\beta}\\
G=&\,\frac{2}{1-4\beta}\,\Big[
\frac{g^2}{4a}\,\frac{(\Delta x)^2}{\Delta t}
(2+\alpha-2\beta)
\Big]^{\frac{1-4\beta}{2(3+\alpha-2\beta)}}
\end{split}
\end{equation}
and express the integral
(\ref{eq:short-time-integral}) as
\begin{equation}
P_{\Delta X}(\Delta x;\Delta t)=\frac{\mathcal{N}\,G}{\sqrt{2\pi\Delta
t\,}}\int_0^\infty
\exp\Big(-\frac{\xi}{2}\psi(u;\eta)\Big)\,\mathrm{d}u
\end{equation}
where
\begin{equation}
\psi(u;\eta)=u^{-\frac{2}{1-4\beta}}+u^{\frac{2(2+\alpha-2\beta)}{1-4\beta}}
-\eta\,u^{\frac{2(1+\alpha-2\beta)}{1-4\beta}}\;.
\label{eq:psi-generic}
\end{equation}
The saddle point approximation is appropriate for large returns,
$|\Delta x|$, which is just the regime of interest. 
The location of the saddle point cannot be found exactly, but it can be expressed as a
series in powers of $\eta$. But as we can see, $\eta$ decreases to
zero when $\Delta x\to\infty$, so the tails of the return distribution
are found both in the limit $\xi\to\infty$ and $\eta\to 0$. Systematic
loop expansion then gives the correction of the order
$O(1/\xi)$. Finally, we get the formula
\begin{equation}
\begin{split}
P_{\Delta X}(\Delta x;\Delta t)=&\,R_N\Big(\frac{|\Delta x|}{\sqrt{\Delta
t}}\Big)^{-\frac{1+\alpha+2\beta}{3+\alpha-2\beta}}\;\times\\
&\times\exp\Bigg[-R_1\,\Big(\frac{|\Delta x|}{\sqrt{\Delta
t}}\Big)^{\frac{2(2+\alpha-2\beta)}{3+\alpha-2\beta}}
+R_2\,\sigma\,\Big(\frac{|\Delta x|}{\sqrt{\Delta
t}}\Big)^{\frac{2(1+\alpha-2\beta)}{3+\alpha-2\beta}}+\\
&+O\Big(\sigma^2\,|\Delta x|^{\frac{2(\alpha-2\beta)}{3+\alpha-2\beta}}\Big)
\Bigg]\times
\\
&\times\Bigg(1+R_3\,\sigma\,\Big(\frac{|\Delta x|}{\sqrt{\Delta
t}}\Big)^{-\frac{2}{3+\alpha-2\beta}}+O\Big(\sigma^2\,|\Delta
x|^{-\frac{4}{3+\alpha-2\beta}}\Big)\Bigg)\times
\\
&\times\Bigg(1+O\Big(|\Delta x|^{-\frac{2(2+\alpha-2\beta)}{3+\alpha-2\beta}}\Big)\Bigg)
\end{split}
\label{eq:distribution-generic}
\end{equation}
where the constants $R_N$, $R_1$, $R_2$, $R_3$ contain the combination
$4a/g^2$ of the remaining parameters
 and depend also on $\alpha$ and $\beta$. 
 The most important aspect of this result is the stretched-exponential
 decay of the tail of the distribution. Therefore, it is fatter than
 the Gaussian, but still not fat enough to explain the power-law
 tails.

There are several special cases to be investigated separately. First,
it is clear that the algebraic substitution which leads to the function
(\ref{eq:psi-generic}) fails if $\beta=1/4$. In this case, exponential
substitution must be used instead, in the form
\begin{equation}
\psi(u;\eta)=\mathrm{e}^{-u}+\mathrm{e}^{(\alpha+3/2)\,u} -
\eta\,\mathrm{e}^{(\alpha+1/2)\,u}  
\label{eq:psi-betaonequarter}
\end{equation}
which leads to a result very similar to
(\ref{eq:distribution-generic}). In fact, it is equivalent to   
 simply setting $\beta=1/4$ in (\ref{eq:distribution-generic}).

More fundamental changes occur in the cases we called previously
``Heston-type'' and ``GARCH-type''. The saddle-point method
is unnecessary here and the integral (\ref{eq:short-time-integral})
can be performed explicitly. Thus, for $1+\alpha-2\beta=0$ we obtain
\begin{equation}
\begin{split}
P_{\Delta X}(\Delta x;\Delta
t)=&\,\frac{2\mathcal{N}}{\big(1-4\beta+4a\,\sigma/g^2\big)
\sqrt{2\pi\,\Delta t\,}}
\Bigg(
{\frac{(\Delta x)^2}{\Delta t}\,\frac{g^2}{4a}}
\Bigg)^{\frac{1}{4}-\beta+\sigma a/g^2}
\times
\\
&\times\,K_{\frac{1}{2}-2\beta+2a\sigma/g^2}\Bigg(\frac{|\Delta x|}{\sqrt{\Delta t}}
\sqrt{\frac{4a}{g^2}\,}
\Bigg)
\end{split}
\end{equation}
with $K_\nu(z)$ the modified Bessel function. From the asymptotic
expansion of the Bessel function we deduce exponential decay of the
tail of the return distribution. This result was already found for the
Heston model itself \cite{dra_yak_02,dra_yak_02b,sil_yak_03}. Now we
can see that this behaviour holds unchanged for the whole
``Heston-type'' class of models.

For $2+\alpha-2\beta=0$
the calculation is even simpler. The result is the algebraic decay of
the return distribution
\begin{equation}
P_{\Delta X}(\Delta x;\Delta
t)=\,\frac{2\mathcal{N}}{\sqrt{2\pi\,\Delta t\,}}
\,\frac{\Gamma\big(-\frac{1}{2}+2\beta+2a/g^2\big)}{
\Big(\frac{(\Delta x)^2}{2\Delta t}+\frac{2a\sigma}{g^2}\Big)
^{-\frac{1}{2}+2\beta+2a/g^2}
}
\end{equation}
valid for entire ``GARCH-type'' class. We have already seen that the
moment method reveals the power-law tail at large times for the
``GARCH'' model, $\alpha=0$, $\beta=1$. Now we can see that, at least
for short time, the power-law tails apply to a wide ensemble of models.

\subsubsection{$\gamma=1$}

The same calculation can be performed for the case $\gamma=1$. For
some values of $\alpha$ and $\beta$ it is sensible to accept also
negative values of the stochastic volatility $S_t$. This is the case,
e.g. of the Stein-Stein model, $\alpha=\beta=0$. However, for generic
values of $\alpha$ and $\beta$ we still need to restrict the
definition range to $S_t>0$. Therefore, we postulate the reflecting
boundary at $S_t=0$ also here. This makes the case $\gamma=1$ entirely
analogous to the previously studied $\gamma=1/2$.

The special cases are $\beta=0$, which requires exponential
substitution, and again the $1+\alpha-2\beta=0$ and
$2+\alpha-2\beta=0$ cases as before. The generic case is characterised
by substitutions
\begin{equation}
\begin{split}
\xi=&\,\Big[\frac{(\Delta x)^2}{\Delta
    t}\Big]^{\frac{2+\alpha-2\beta}{4+\alpha-2\beta}}
\;\Big[\frac{g^2}{4a}(2+\alpha-2\beta)\Big]^{-\frac{2}{4+\alpha-2\beta}}\\
\eta=&\,\sigma\,\Big[\frac{g^2}{4a}\,\frac{(\Delta x)^2}{\Delta
    t}\Big]^{-\frac{1}{4+\alpha-2\beta}}\;\;
\frac{(2+\alpha-2\beta)^{\frac{3+\alpha-2\beta}{4+\alpha-2\beta}}}{1+\alpha-2\beta}\\
G=&\,\frac{1}{2\beta}\,\Big[
\frac{g^2}{4a}\,\frac{(\Delta x)^2}{\Delta t}
(2+\alpha-2\beta)
\Big]^{-\frac{2\beta}{4+\alpha-2\beta}}
\end{split}
\end{equation}
The integral to perform is
\begin{equation}
P_{\Delta X}(\Delta x;\Delta t)=\frac{\mathcal{N}\,G}{\sqrt{2\pi\Delta
t\,}}\int_0^\infty
\exp\Big(-\frac{\xi}{2}\psi(u;\eta)\Big)\,\mathrm{d}u
\end{equation}
where 
\begin{equation}
\psi(u;\eta)=u^{\frac{1}{\beta}}+u^{-\frac{2+\alpha-2\beta}{2\beta}}
-\eta\,u^{-\frac{1+\alpha-2\beta}{2\beta}}\;.
\label{eq:psi-generic-gammaone}
\end{equation}

Then, the return distribution for $\Delta x\to\infty $ is 
\begin{equation}
\begin{split}
P_{\Delta X}(\Delta x;\Delta t)=&\,R_N\Big(\frac{|\Delta x|}{\sqrt{\Delta
t}}\Big)^{-\frac{2+\alpha+2\beta}{4+\alpha-2\beta}}\;\times\\
&\times\exp\Bigg[-R_1\,\Big(\frac{|\Delta x|}{\sqrt{\Delta
t}}\Big)^{\frac{2(2+\alpha-2\beta)}{4+\alpha-2\beta}}
+R_2\,\sigma\,\Big(\frac{|\Delta x|}{\sqrt{\Delta
t}}\Big)^{\frac{2(1+\alpha-2\beta)}{4+\alpha-2\beta}}+\\
&+O\Big(\sigma^2\,|\Delta x|^{\frac{2(\alpha-2\beta)}{4+\alpha-2\beta}}\Big)
\Bigg]\times
\\
&\times\Bigg(1+R_3\,\sigma\,\Big(\frac{|\Delta x|}{\sqrt{\Delta
t}}\Big)^{-\frac{2}{4+\alpha-2\beta}}+O\Big(\sigma^2\,|\Delta
x|^{-\frac{4}{4+\alpha-2\beta}}\Big)\Bigg)\times
\\
&\times\Bigg(1+O\Big(|\Delta x|^{-\frac{2(2+\alpha-2\beta)}{4+\alpha-2\beta}}\Big)\Bigg)
\end{split}
\label{eq:distribution-generic-gammaone}
\end{equation}
with, again, some constants  $R_N$, $R_1$, $R_2$, $R_3$, in full
analogy with (\ref{eq:distribution-generic}).  

Explicit calculation shows that the case $\beta=0$, treated by
exponential substitution and saddle point method, gives again the same
result as direct substitution $\beta=0$ in (\ref{eq:distribution-generic-gammaone}).

The special cases are treated similarly. For $1+\alpha-2\beta=0$ we
have the following substitutions
\begin{equation}
\begin{split}
\xi=&\,\Big[\frac{(\Delta x)^2}{\Delta
    t}\Big]^{\frac{1}{3}}
\;\Big(\frac{g^2}{4a}\Big)^{-\frac{2}{3}}\\
G=&\,\frac{1}{-2\beta+2a\sigma/g^2}\,\Big[
\frac{g^2}{4a}\,\frac{(\Delta x)^2}{\Delta t}
\Big]^{-\frac{2\beta}{3}+\frac{2a\sigma}{3g^2}}
\end{split}
\end{equation}
and the function within the exponential is now
\begin{equation}
\psi(u)=u^{-2/(-2\beta+2a\sigma/g^2)}+u^{1/(-2\beta+2a\sigma/g^2)}
\;.
\label{eq:psi-hestontype-gammaone}
\end{equation}
The resulting return distribution can be expressed by generalised
hypergeometric functions, but it is more transparent to perform the
saddle-point calculations, which give the tails
\begin{equation}
\begin{split}
P_{\Delta X}(\Delta x;\Delta t)=&\,R_N
\Big(\frac{|\Delta x|}{\sqrt{\Delta
t}}\Big)^{-\frac{1}{3}-\frac{4}{3}\beta+\frac{4a\sigma}{3g^2}}\;
\exp\Big[-R_1\,\Big(\frac{|\Delta x|}{\sqrt{\Delta
t}}\Big)^{\frac{2}{3}}\Big]
\times
\\
&\times\Bigg(1+O\Big(|\Delta x|^{-\frac{2}{3}}\Big)\Bigg)
\end{split}
\label{eq:distribution-hestontype-gammaone}
\end{equation}

The remaining special case is  $2+\alpha-2\beta=0$. The integral to
perform has the following form
\begin{equation}
P_{\Delta X}(\Delta x;\Delta t)=\frac{\mathcal{N}}{\sqrt{2\pi\Delta
    t}}
\int_0^\infty s^{-1-2\beta-2a/g^2}\;\exp\Big(
-\frac{(\Delta x)^2}{2\Delta t\,s^2}-\frac{2a\sigma}{g^2\,s}\Big)\;.
\label{eq:integral-garchtype-gammaone}
\end{equation}
Here, the
saddle-point method is of no use, because the corresponding function
$\psi(u)$, if used, would have no local extreme. Instead, we can use
a simple trick to estimate the behaviour of the tail, at $|\Delta
x|\to\infty$. The exponential function within the integrand is close 
to $1$ for $s\gtrsim s^*\simeq |\Delta x|/\sqrt{\Delta t}$ and very
small for $s\lesssim s^*$. More precisely, this 
behaviour holds only for $|\Delta
x|/\sqrt{\Delta t}\gtrsim 2a\sigma/g^2$, but for the discussion of the
tail it is quite sufficient. Thus, the integral
(\ref{eq:integral-garchtype-gammaone}) can be estimated by integrating
only the algebraic factor from $s^*$ to infinity. Finally, we obtain
the power-law tail of the return distribution in the form
\begin{equation}
P_{\Delta X}(\Delta x;\Delta t)\sim
\Big(\frac{|\Delta x|}{\sqrt{\Delta t}}\Big)^{-2\beta-2a/g^2}\;.
\end{equation}

As a summary, we can see that the short-time return distribution has
generically a tail in the form of a stretched exponential. This
becomes simple exponential in the case of $\gamma=1/2$ and Heston-type
combination of parameters, $1+\alpha-2\beta=0$. The exceptional case
is the family of ``GARCH''-type models, characterised by the
combination  $2+\alpha-2\beta=0$. In that case, we found that both for
$\gamma=1/2$ and for $\gamma=1$ the tail of the return distribution has
a power-law form.

As a separate calculation, we show how the same procedure works for
the expOU model. We have to perform the integral
\begin{equation}
P_{\Delta X}(\Delta x;\Delta t)=\sqrt{\frac{a}{\pi^2\Delta
    t\,g^2}}
\int_0^\infty 
\mathrm{e}^{-a\,\psi(u)/(2g^2)}\,\mathrm{d}u
\label{eq:integral-expou}
\end{equation}
where
\begin{equation}
\psi(u)=2\xi\,u+\big(\ln u\big)^2
\end{equation}
and
\begin{equation}
\xi=\frac{g^2\,(\Delta x)^2}{2a\,\Delta t}\;.
\end{equation}
The saddle point $u^*$ can be expressed using the Lambert function
$W_\mathrm{L}(y)$ defined by the equation
$W_\mathrm{L}(y)\,\mathrm{e}^{W_\mathrm{L}(y)}=y$. Finally, we find
that 
\begin{equation}
\psi(u^*)=\big(1+W_\mathrm{L}(\xi)\big)^2-1
\end{equation}
and within the saddle-point approximation
\begin{equation}
P_{\Delta X}(\Delta x;\Delta t)\propto
\exp\Big\{-\frac{a}{2g^2}\Big[\Big(1+W_\mathrm{L}\Big(\frac{g^2\,(\Delta x)^2}{2a\,\Delta t}\Big)\Big)^2-1\Big]\Big\}\;.
\label{eq:distribution-expou-saddle}
\end{equation}

The tails of the return distributions, $\Delta x\to\infty$, correspond
again to the limit $\xi\to\infty$. However, the situation is less
transparent here, because the Lambert function does not have a simple
asymptotic behaviour. Naively, one could use the asymptotics
\begin{equation}
W_\mathrm{L}(y)\simeq \ln y,\qquad y\to\infty
\label{eq:lambert-asympt}
\end{equation}
but it holds well only for extremely large $y$. There is a much better
approximate formula
\begin{equation}
W_\mathrm{L}(y)\simeq \ln y+\Big(\frac{1}{1+\ln y}-1\Big)\,\ln\ln y,\qquad y\to\infty
\end{equation}
which is reasonably accurate already for $y \gtrsim  3$. 
The farthest tail of the return distribution, using
(\ref{eq:distribution-expou-saddle}) and (\ref{eq:lambert-asympt}), has
the form 
\begin{equation}
P_{\Delta X}(\Delta x)\sim (\Delta x)^{-(2a/g^2)\ln \Delta x}
\end{equation}
which is steeper than any power but fatter than any stretched
exponential. Therefore, the expOU model lies somewhere between the ``GARCH''
model and the other models exhibiting stretched exponential tails. In
this sense it is very promising, as the true power-law tails are
hardly detectable in real data, but the behaviour very close to power
law is well documented. The exponential Ornstein-Uhlenbeck process may
serve as a convenient tool in this situation.

\section{Conclusions}

We showed that many properties of wide class of stochastic volatility
models can be deduced quite generally without full explicit solution
of the Fokker-Planck equation for the set of coupled stochastic
differential equations. The class characterised by algebraic
coefficients was solved systematically in the long-time and short-time
regimes. Full solution for the stationary autocorrelation functions
was found in two special cases, while generic features of the
autocorrelation function are found generally.

Specifically, the method of moments was shown to reveal the asymptotic
distribution of return and volatility. The return distribution is
asymptotically Gaussian, except for the ``GARCH'' model,
where power-law tail develops. This tail in returns originates from
analogous power-law tail in the distribution of volatility. The
empirical findings \cite{li_go_ci_me_pe_sta_99,bou_pot_03} suggest
that the central part of the volatility distribution is close to
log-normal, while the tail is fatter, perhaps indeed a
power-law. This makes the ``GARCH'' stochastic volatility model a good
candidate for phenomenological modelling and risk assessment. Saying
``phenomenological'' we admit and stress that the model does not say
anything of the real mechanism how the price fluctuations are produced,
neither it explains why just the chosen combination of the parameters
$\alpha$, $\beta$ and $\gamma$ is superior to others.

While the price process is not Markovian any more, the compound
price-volatility process does have the Markov property by
definition. It results in exponential decay of volatility
autocorrelation. Multiple time scales found in the expOU model may be
used as explanation for the slow decay of empirical autocorrelations
\cite{mas_per_05,lebaron_01}. However, to balance the opinions, we
should also note that the approach of \cite{lebaron_01} was severely
criticised in \cite{sta_ple_01}.

The fat tails in return distribution, which is the most notorious
stylised fact of empirical econophysics, are in fact a short-time
property. At long time distances, the return distribution approaches
to a Gaussian, as if the price process was a Brownian motion. This
lead us to the study of short-time properties of the
stochastic-volatility models, which at long times exhibit
undifferentiated (with exception of the ``GARCH'' class) Gaussian
behaviour. We examined the return distributions for general values of
the parameters $\alpha$, $\beta$ and $\gamma$ and we found that
stretched exponential is the generic form of the distribution. As a
special case, we find exponential distribution in the Heston-type
class, $1+\alpha-2\beta=0$. One exception is the ``GARCH''-type class, ($2+\alpha-2\beta=0$)
where the distribution has power-law tail. Of course, this is not
surprising, as we already know that the  power-law tail persists for
all times. The intermediate class is represented by the exponential
Ornstein-Uhlenbeck process, where the tail is thinner than a
power-law, but fatter than stretched exponential. 

To sum up, we examined systematically whole class of stochastic
volatility models and found explicitly their asymptotic and short time
properties, as well as the volatility autocorrelation. Power-law tails
are reproduces in the family of ``GARCH''-type models, while generic
form of the short-time return distribution is stretched exponential.

\section*{Acknowledgements} 

This work was carried out within the project AV0Z10100520 of the Academy 
of Sciences of the Czech Republic and was  
supported by the M\v SMT of the Czech Republic, grant no. 
OC09078 and by the Research Program CTS MSM 0021620845.


\begin{thebibliography}{99}
%Bib. item no.: 1
%Record: 2644
\bibitem{lewis_00}
A. L. Lewis,
{\it Option Valuation under Stochastic Volatility}
 (Finance Press, Newport Beach, 2000).

%Bib. item no.: 2
%Record: 2635
\bibitem{cis_fer_mon_nic_07}
E. Cisana, L. Fermi, G. Montagna, and O. Nicrosini,
arXiv:0709.0810
 (2007).

%Bib. item no.: 3
%Record: 2455
\bibitem{bou_mat_pot_01}
J.-P. Bouchaud, A. Matacz, and M. Potters,
Phys. Rev. Lett.
 { 87}
 (2001)
 228701.

%Bib. item no.: 4
%Record: 2472
\bibitem{mas_per_02}
J. Perell\'o and J. Masoliver,
Int. J. Theor. Appl. Finance
 { 5}
 (2002)
 541.

%Bib. item no.: 5
%Record: 2457
\bibitem{per_mas_bou_03}
J. Perell\'o, J. Masoliver, and J.-P. Bouchaud,
Applied Mathematical Finance
 { 11}
 (2004)
 27.

%Bib. item no.: 6
%Record: 2458
\bibitem{per_mas_03}
J. Perell\'o and J. Masoliver,
Phys. Rev. E
 { 67}
 (2003)
 037102.

%Bib. item no.: 7
%Record: 2463
\bibitem{per_mas_ane_04}
J. Perell\'o, J. Masoliver, and N. Anento,
Physica A
 { 344}
 (2004)
 134.

%Bib. item no.: 8
%Record: 2845
\bibitem{bar_she_01}
O. E.  Barndorff-Nielsen and N.  Shephard,
J. Royal Stat. Soc. B
 { 63}
 (2001)
 167.

%Bib. item no.: 9
%Record: 2632
\bibitem{scott_87}
L. O. Scott,
Journal of Financial  and Quantitative Analysis
 { 22}
 (1987)
 419.

%Bib. item no.: 10
%Record: 2844
\bibitem{lebaron_01}
B. LeBaron,
Quant. Finance
 { 1}
 (2001)
 621.

%Bib. item no.: 11
%Record: 2117
\bibitem{mas_per_05}
J. Masoliver and J. Perell\'o,
Quant. Finance
 { 6}
 (2006)
 423.

%Bib. item no.: 12
%Record: 2459
\bibitem{eis_per_mas_06}
Z. Eisler, J. Perell\'o and J. Masoliver,
Phys. Rev. E
 { 76}
 (2007)
 056105.

%Bib. item no.: 13
%Record: 2630
\bibitem{per_sir_mas_08}
J. Perell\'o, R. Sircar, and J. Masoliver,
J. Stat. Mech.
 (2008)
 P06010.

%Bib. item no.: 14
%Record: 2634
\bibitem{bor_caz_mon_nic_08}
G. Bormetti, V. Cazzola, G. Montagna, and O. Nicrosini,
J. Stat. Mech.
 (2008)
 P11013.

%Bib. item no.: 15
%Record: 2487
\bibitem{mas_per_07}
J. Masoliver and J. Perell\'o,
Phys. Rev. E
 { 75}
 (2007)
 046110.

%Bib. item no.: 16
%Record: 2814
\bibitem{bar_ciu_spa_93}
P. Barrera, S. Ciuchi, and B. Spagnolo,
J. Phys. A: Math. Gen.
 { 26}
 (1993)
 L559.

%Bib. item no.: 17
%Record: 2813
\bibitem{ciu_pas_spa_93}
S. Ciuchi, F. de Pasquale, and B. Spagnolo,
Phys. Rev. E
 { 47}
 (1993)
 3915.

%Bib. item no.: 18
%Record: 2486
\bibitem{vasicek_77}
O. Vasicek,
Journal of Financial Economics
 { 5}
 (1977)
 177.

%Bib. item no.: 19
%Record: 2485
\bibitem{cox_ing_ros_85}
J. C. Cox, J. E. Ingersoll Jr., S. A. Ross,
Econometrica
 { 53}
 (1985)
 385.

%Bib. item no.: 20
%Record: 2478
\bibitem{feller_51}
W. Feller,
Ann. Math.
 { 54}
 (1951)
 173.

%Bib. item no.: 21
%Record: 2491
\bibitem{hul_whi_90}
J. Hull and A. White,
The Review of Financial Studies
 { 3}
 (1990)
 573.

%Bib. item no.: 22
%Record: 2633
\bibitem{hul_whi_87}
J. Hull and A. White,
Journal of Finance
 { 42}
 (1987)
 281.

%Bib. item no.: 23
%Record: 2480
\bibitem{ste_ste_91}
E. M. Stein and J. C. Stein,
The Review of Financial Studies
 { 4}
 (1991)
 727.

%Bib. item no.: 24
%Record: 2498
\bibitem{bal_rom_94}
C. A. Ball and  A. Roma,
Journal of Financial and Quantitative Analysis
 { 29}
 (1994)
 589.

%Bib. item no.: 25
%Record: 2479
\bibitem{sch_zhu_99}
R. Sch\"obel and J. Zhu,
European Finance Review
 { 3}
 (1999)
 23.

%Bib. item no.: 26
%Record: 2481
\bibitem{heston_93}
S. L. Heston,
The Review of Financial Studies
 { 6}
 (1993)
 327.

%Bib. item no.: 27
%Record: 1880
\bibitem{dra_yak_02}
A. A. Dr\u{a}gulescu and V. M. Yakovenko,
Quant. Finance
 { 2}
 (2002)
 443.

%Bib. item no.: 28
%Record: 2044
\bibitem{dra_yak_02b}
A. A. Dr\u{a}gulescu and V. M. Yakovenko,
Quant. Finance
 { 2}
 (2002)
 443.

%Bib. item no.: 29
%Record: 2043
\bibitem{sil_yak_03}
A. C. Silva and V. M. Yakovenko,
Physica A
 { 324}
 (2003)
 303.

%Bib. item no.: 30
%Record: 2041
\bibitem{sil_pra_yak_04}
A. C. Silva, R. E. Prange, and V. M. Yakovenko,
Physica A
 { 344}
 (2004)
 227.

%Bib. item no.: 31
%Record: 2042
\bibitem{vic_tol_lei_cat_04}
R. Vicente, C. M. de Toledo, V. B. P. Leite, and N. Caticha,
Physica A
 { 361}
 (2006)
 272.

%Bib. item no.: 32
%Record: 2474
\bibitem{sil_yak_06}
A. C. Silva and V. M. Yakovenko,
Physica A
 { 382}
 (2007)
 278.

%Bib. item no.: 33
%Record: 2631
\bibitem{jiz_kle_hae_07}
P. Jizba, H. Kleinert, and P. Haener,
Physica A
 { 388}
 (2009)
 3503.

%Bib. item no.: 34
%Record: 2464
\bibitem{pal_per_mon_mas_04}
L. Palatella, J. Perell\'o, M. Montero, and J. Masoliver,
Eur. Phys. J. B
 { 38}
 (2004)
 671.

%Bib. item no.: 35
%Record: 2488
\bibitem{ric_sab_04}
P. Richmond and L. Sabatelli,
Physica A
 { 336}
 (2004)
 27.

%Bib. item no.: 36
%Record: 2002
\bibitem{ant_rie_05}
C. Anteneodo and R. Riera,
Phys. Rev. E
 { 72}
 (2005)
 026106.

%Bib. item no.: 37
%Record: 2462
\bibitem{labordere_05}
P. Henry-Labord\`ere,
Quant. Finance
 { 7}
 (2007)
 525.

%Bib. item no.: 38
%Record: 2490
\bibitem{labordere_05a}
P. Henry-Labord\`ere,
cond-mat/0504317.

%Bib. item no.: 39
%Record: 2461
\bibitem{bon_val_spa_06}
G. Bonanno, D. Valenti, and B. Spagnolo,
Phys. Rev. E
 { 75}
 (2007)
 016106.

%Bib. item no.: 40
%Record: 2460
\bibitem{bon_val_spa_05}
G. Bonanno, D. Valenti, and B. Spagnolo,
Eur. Phys. J. B
 { 53}
 (2006)
 405.

%Bib. item no.: 41
%Record: 2477
\bibitem{val_spa_bon_06}
D. Valenti, B. Spagnolo, and G. Bonanno,
Physica A
 { 382}
 (2007)
 311.

%Bib. item no.: 42
%Record: 2489a
\bibitem{villaroel_06}
J. Villarroel,
Physica A
 { 382}
 (2007)
 321.

%Bib. item no.: 43
%Record: 2816
\bibitem{spa_val_08}
B. Spagnolo and D. Valenti,
Int. J. Bif. Chaos
 { 18}
 (2008)
 2775.

%Bib. item no.: 44
%Record: 1225
\bibitem{go_ple_am_me_sta_99}
P. Gopikrishnan, V. Plerou, L. A. N. Amaral, M. Meyer, and H. E. Stanley,
Phys. Rev. E
 { 60}
 (1999)
 5305.

%Bib. item no.: 45
%Record: 1527
\bibitem{ple_gop_am_mey_sta_99}
V. Plerou, P. Gopikrishnan, L. A. N. Amaral, M. Meyer, and H. E. Stanley,
Phys. Rev. E
 { 60}
 (1999)
 6519.

%Bib. item no.: 46
%Record: 631
\bibitem{ci_li_me_pe_sta_97}
P.~Cizeau, Y.~Liu, M.~Meyer, C.-K.~Peng, H.~E.~Stanley,
Physica A
 { 245}
 (1997)
 441.

%Bib. item no.: 47
%Record: 1312
\bibitem{li_go_ci_me_pe_sta_99}
Y. Liu, P. Gopikrishnan, P. Cizeau, M. Meyer, C.-K. Peng, and H. E. Stanley,
Phys. Rev. E
 { 60}
 (1999)
 1390.

%Bib. item no.: 48
%Record: 1409b
\bibitem{bou_pot_03}
J.-P. Bouchaud and M. Potters,
{\it Theory of Financial Risk  and Derivative Pricing}
 (Cambridge University Press, Cambridge, 2003).

%Bib. item no.: 49
%Record: 2739
\bibitem{mic_bon_lil_man_02}
S. Miccich\`e, G. Bonanno, F. Lillo, and R. N. Mantegna,
Physica A
 { 314}
 (2002)
 756.

%Bib. item no.: 50
%Record: 2815
\bibitem{suz_kan_tak_82}
M. Suzuki, K. Kaneko, and S. Takesue,
Prog. Theor. Phys.
 { 67}
 (1982)
 1756.

%Bib. item no.: 51
%Record: 2482
\bibitem{bir_ros_07}
T. S. Bir\'o and R. Rosenfeld,
Physica A
 { 387}
 (2008)
 1603.

%Bib. item no.: 52
%Record: 2040
\bibitem{sta_ple_01}
H. E. Stanley and V. Plerou,
Quant. Finance
 { 1}
 (2001)
 563.

%
%
\end{thebibliography}
\end{document}